\documentclass[twocolumn,preprintnumbers,secnumarabic,amsmath,amssymb,nofootinbib,floatfix]{revtex4}
\usepackage{varioref,exscale,latexsym,amsmath,amssymb}
\usepackage{graphicx}
\usepackage{graphicx}% Include figure files
\usepackage{dcolumn}% Align table columns on decimal point
\usepackage{bm}% bold math
\usepackage{slashed}

\def\beq{\begin{equation}}

\def\eeq{\end{equation}}

\def\beqa{\begin{eqnarray}}

\def\eeqa{\end{eqnarray}}

\begin{document}
\title{{\bf Phenomenology of the Equivalence Principle with Light Scalars \\}}
\medskip
\author{Thibault Damour$^{a}$}
\author{ John F. Donoghue$^{a,b}$}
\affiliation{${}^a$Institut des Hautes \'{E}tudes Scientifiques \\
Bures sur Yvette, F-91440, France\\
and \\
${}^b$Department of Physics\\
University of Massachusetts\\
Amherst, MA  01003, USA
}

\begin{abstract}
Light scalar particles with couplings of sub-gravitational strength, which can generically be called `dilatons', can produce violations of the equivalence principle. However, in order to understand experimental sensitivities one must know the coupling of these scalars to atomic systems. We report here on a study of the required couplings. We give a general Lagrangian with five independent dilaton parameters and calculate the ``dilaton charge'' of atomic systems for each of these. Two combinations are particularly important. One is due to the variations in the nuclear binding energy, with a sensitivity scaling with the atomic number as $A^{-1/3}$. The other is due to electromagnetism. We compare limits on the dilaton parameters from existing experiments.
\end{abstract}
\maketitle

The Equivalence Principle (EP) is one of the most exquisitely tested features in science, with
a present sensitivity of
\begin{equation}
\frac{\Delta a}{a} \sim 10^{-13}
\end{equation}
for comparisons of the acceleration of two test masses in a gravitational field \cite{Schlamminger:2007ht, lunarlaser}.
Further tests of this principle remain important {\em because} of this high precision - the
tests can be sensitive to very small effects that other experiments might not notice \cite{damour,Turyshev:2009ir}.
Indeed, new space-based experiments such as MICROSCOPE \cite{microscope}, the Galileo Galilei project \cite{gg} and
STEP \cite{step} as well as
innovative experiments with cold atoms \cite{cold atoms, mueller} and sub-orbital rockets \cite{POEM} are being planned that could push the sensitivity several
orders of magnitude further.

In comparing experiments, it is common to invoke a hypothetical model with the exchange of a light vector particle coupled to baryon number. This is convenient because it is easy to know the baryon number of a given material, and hence the coupling of this vector particle to atomic systems is known. However this is an unlikely physical situation, as very light and very weakly coupled vectors are rare in modern theories. A more plausible candidate as a source of EP violation is a light\footnote{We will generally assume in the following that the scalar
field we consider is essentially massless on the scales that we discuss, although is it simple to modify this assumption.} scalar field with a coupling to matter
that is weaker than gravitational strength. We will refer to these generically as `dilatons', although they may have origins other than
string theory or models involving dilation symmetry \cite{Damour:1994ya, runaway, Damour:1992we, Kaplan}. Being Lorentz scalars, dilatons will couple to Lorentz invariant combinations of the Standard Model fields, the quarks, gluons and photons. Significant theoretical effort is needed to translate these fundamental interactions into the required couplings of the atomic systems used in experiments. In \cite{damourdonoghueEP} we provide an extensive study of the theoretical ingredients to this program, and in this paper we summarize the results which are most useful to experiments.

The important physics is readily understood, even if it is somewhat more difficult to calculate. Most of the mass of atomic systems comes from the gluonic QCD interactions, which sets the scale for the strong interactions. However, atomic masses also have smaller dependencies on the masses of the light quarks and on the electric charge. This can occur either in the masses of the individual nucleons or in the binding energy of the nucleus. General relativity couples universally to all forms of energy, but dilaton couplings can and will vary among the different ingredients to the total atomic mass. We find particular sensitivity in the binding energy coming from the quark mass dependence, and this is the most novel aspect of our analysis\footnote{Damour \cite{damour}  and Dent  \cite{dent} have highlighted this
need for the study of the nuclear binding energies.}. As described below, in addition to the well known dependence on electromagnetism, the binding energy has an enhanced dependence on the quark masses because of the importance of the pion in nuclear binding and the strong dependence of the pion mass on the light quark masses. Although we provide a description of all effects, we will also provide a simpler parameterization of the dominant hadronic binding sensitivity, proportional to $A^{-1/3}$, and the electromagnetic sensitivy, proportional to $Z(Z-1)/A^{4/3}$ .

A massless dilaton $\phi$ modifies the Newtonian interaction between a mass $A$ and a mass $B$, into the form
(see, e.g. \cite{Damour:1992we})
\begin{equation}
V=-G\frac{m_A m_B}{r_{AB}}(1+\alpha_A \alpha_B).
\end{equation}
If the dilaton mass is important the second term includes an extra exponential factor $\exp(-m_\phi r_{AB})$.
In this interaction potential, the scalar coupling to matter is measured by the dimensionless factors, $\alpha_{A,B}$. To calculate these, we add the dilaton Lagrangian to that of the Standard Model and use the result to evaluate the total mass of the atomic system. The $\alpha_A$ coupling is then found via
\begin{equation}
\label{alphaphi}
\alpha_A = \frac{1}{\kappa m_A}\frac{\partial [\kappa m_A(\phi)] }{\partial [\kappa\phi]} .
\end{equation}
Here,  $\kappa \equiv \sqrt{ 4\pi G}$ is the inverse of the Planck mass\footnote{We use units such that $c=1=\hbar$.} so that the products
$\kappa m_A$ and $\kappa\phi$ are dimensionless. This ensures that this definition of $\alpha_A$ is valid in any choice of units.
In terms of the $ \alpha_A$'s, the violation of the (weak) EP, i.e. the fractional difference between the accelerations of two bodies $A$ and $B$
falling in the gravitational field generated by an external body $E$, reads
\begin{eqnarray}
\label{da/a}
\left( \frac{\Delta a}{a} \right)_{AB} &\equiv& 2\frac{a_A-a_B}{a_A+a_B}
= \frac{(\alpha_A - \alpha_B)\alpha_E}{{1+ \frac12(\alpha_A + \alpha_B)\alpha_E}} \nonumber \\
&\simeq& (\alpha_A- \alpha_B)\alpha_E.
\end{eqnarray}
In the last (approximate) equation we have assumed that the $\alpha$'s are small.

We couple the dilaton to the light fields of the Standard Model. The heavy quarks and weak gauge bosons are assumed to be integrated out and we assume that the dilaton theory has been matched to the light fields below the scale of the heavy quarks. This procedure introduces {\em five} dimensionless dilaton-coupling
parameters, $d_e, d_g$ for the couplings to the electromagnetic and gluonic field-strength terms, and $d_{m_e}, d_{m_u}, d_{m_d}$
for the couplings to the fermionic mass terms. [We are using here the fact that a $\phi-$dependent coupling to the kinetic term
of a fermion, $f(\phi) {\bar \psi} i {\slashed{D}} \psi$, can be absorbed in a suitable $\phi-$dependent rescaling of $\psi$.] We add these interactions to that of the Standard Model,  ${\cal L }= {\cal L }_{SM} + {\cal L }_{\rm int}$ and normalize
these five dimensionless dilaton parameters $d_e, d_g, d_{m_e}, d_{m_u}, d_{m_d}$ so that they
correspond (when considering the linear couplings to $\phi$) to
\begin{eqnarray}
\label{Lint}
{\cal L}_{{\rm int} } &=&  \kappa \phi \left[ + \frac{d_e}{4e^2} F_{\mu\nu}F^{\mu\nu}
-\frac{d_g\beta_3}{2g_3} F^A_{\mu\nu}F^{A\mu\nu} \right. \nonumber \\  &-& \left.  \sum_{i=e,u,d} (d_{m_i}+\gamma_{m_i}d_g) m_i{\bar \psi}_i\psi_i
\right]
\end{eqnarray}
These are chosen to correspond to renormalization group invariants \cite{Kaplan, damourdonoghueEP}. For the quark mass terms, physics tells us that it is preferable to work with the symmetric and antisymmetric combinations
\begin{equation}
\hat{m}=\frac{1}{2}(m_d+m_u) \,, \qquad
\delta m=(m_d-m_u)
\end{equation}
so that we define dilaton parameters ${d}_{\hat m},~~d_{\delta m}$ conjugate to
these combinations
\begin{equation}
{\cal L}_{\rm int} = .... -\kappa \phi \left[{d}_{\hat m} \hat{m}(\bar{d}d +  \bar{u}u)  + \frac{d_{\delta m}}{2}\delta m (\bar{d}d -  \bar{u}u)\right]
\end{equation}
and use these parameters instead of $d_{m_u},~~d_{m_d}$.

We need to calculate the effect of these couplings for atoms. In order to do this we transform the scalar field dependence into an implicit dependence on the parameters of the Standard Model, and use the decades of research connecting the Standard Model to observable physics. For example, we follow the pioneering method of \cite{Damour:1994ya} for the electromagnetic coupling
\begin{equation}
\label{EM}
{\cal L}_{EM} = -\frac{1 - d_e\kappa \phi}{4e^2} F_{\mu\nu}F^{\mu\nu}
\simeq -\frac{1 }{4 (1 + d_e\kappa \phi)e^2} F_{\mu\nu}F^{\mu\nu}
\end{equation}
where the last equality is valid at the linear level in $\kappa \phi$ (which is the level at which we
define the dilaton couplings here).
As we work with a rescaled  electromagnetic field ($ A^{\rm here} = e A^{\rm usual}$), the only location where the electric charge occurs in the Lagrangian is
the one explicitly shown above. This allows the dilaton field to be absorbed into the following $\phi$ dependence of the fine-structure constant $\alpha=e^2/(4 \pi)$
\begin{equation}
\alpha(\phi) =   (1 + d_e\kappa\phi ) \alpha  ~~.
\end{equation}
and allows us to obtain the dilaton coupling from the dependence of the masses on the electric charge
\begin{equation}
\alpha_A^{(d_e)}  = \frac{d_e}{m_A}\alpha \frac{\partial m_A }{\partial \alpha}~~~.
\end{equation}
Likewise the fermion mass terms are normalized such that we can define a field-dependent mass
\begin{equation}
m_i(\phi) = (1+d_{m_i}\kappa\phi)m_i~,~~~~i=u,d,e
\end{equation}
and we obtain the dilaton coupling from the dependence of the atomic masses on the fermion masses. Finally, the terms proportional to the gluonic coupling $d_g$ are normalized so that they are proportional to the QCD trace of the energy momentum tensor \cite{Kaplan, damourdonoghueEP}, which allows the matrix element to be readily calculated. This turns the parameter $d_g$ into one that measures the $\phi$ sensitivity of the QCD mass scale, say  $\Lambda_3$. We then have
 \begin{equation}
\frac{\partial \ln \Lambda_3}{\partial [ \kappa \phi]}=  d_g \, ,~~~~
 \frac{\partial \ln m_i(\Lambda_3)}{\partial [ \kappa \phi]}= d_{m_i}\, .
\end{equation}
Variation of dimensionless ratios such as $\ln (m_i/\Lambda_3)$ then involve differences of the coupling parameters $d_m-d_g$.

The structure of the coupling to individual nucleons is reasonably well known.
The quark mass contributions are known from the nucleon sigma term \cite{sigma} and from the baryon mass splittings, and electromagnetic contributions to the masses have been estimated \cite{masses}.

Nuclear binding is well parameterized by the semi-empirical mass formula\footnote{For simplicity in this Letter we drop the pairing interaction as it does not play a significant role in our analysis},
\begin{equation}
 E^{\rm bind} =- a_v A +a_s A^{2/3} +a_a \frac{(A-2Z)^2}{A} + a_c \frac{Z(Z-1)}{A^{1/3}}  \, .
\end{equation}
Typical fit values for these parameters are \cite{massformula} $a_v =16~ {\rm MeV}, a_s=17 ~{\rm MeV}, a_a= 23 ~{\rm MeV}, a_p= 12 ~{\rm MeV},
a_c = 0.717 ~{\rm MeV} $.
The Coulomb term $a_c$ is linear in the electromagnetic fine structure constant and therefore directly yields the electromagnetic coupling that we need. For the hadronic component, we use our previous work on the quark-mass dependence of nuclear binding \cite{chiral}. The primary physics here is that the central nuclear potential has an important component from the exchange of two pions. Because of the nature of chiral symmetry in QCD, the pion mass-squared is directly proportional to the average light quark mass, $m_\pi^2\sim \hat{m}$.
In the overall binding energy, there is a partial cancelation between the attractive two pion component, which is particularly sensitive to the pion mass, and the repulsive short range potential, which is less sensitive. The variation with pion mass then is stronger than simple expectations. We have estimated the variation of both the central terms $a_v,~a_s$ and the asymmetry energy $a_a$ and find the following contribution to $\alpha_A$
\begin{eqnarray}
\label{baralphabind}
\bar{\alpha}_A^{{\rm bind}} &=& (d_{\hat{m}} -d_g)  F_A\times \left[ 0.045 -\frac{0.036}{A^{1/3}}\right.   \nonumber \\
&-& \left. 0.020 \frac{(A-2Z)^2}{A^2}  - 1.42 \times 10^{-4} \, \frac{Z(Z-1)}{A^{4/3}} \right].
\end{eqnarray}
Here we use the notation
\begin{equation}
\label{FA}
F_A \equiv \frac{A \, m_{\rm amu}}{m_A}
\end{equation}
where we take $m_{\rm amu}=931$~MeV as the nucleon mass with the average binding energy, 8~MeV, subtracted. This factor is very close to unity throughout the periodic table.
We estimate a $30\%$ uncertainty in these numbers, and note that our formalism in \cite{damourdonoghueEP} is most reliable for heavier elements.

In order to highlight the difference between various materials we define
\begin{equation}
{\alpha}_A = d_g +  \bar{\alpha}_A
\end{equation}
The overall common coupling $d_g$ does not violate the EP, while the EP variation is contained in $\bar{\alpha}$. The result of our calculation can be summarized in four `dilaton charges' $Q_{\hat m},~Q_{\delta m},~ Q_{m_e},~Q_e$. Each charge gives the strength of the EP-violating coupling
corresponding to a given dilaton parameter for an atom of charge $Z$ and atomic number $A$, namely
\begin{eqnarray}
\label{baralpha4}
\bar{\alpha}_A &=&  \left[ (d_{\hat m} - d_g) Q_{\hat m} +(d_{\delta m} -d_g) Q_{\delta m}  \right. \nonumber \\
&~&\left. ~+ (d_{m_e}-d_g) Q_{m_e} + d_e Q_e \right]_A ~~.
\end{eqnarray}
The four dilaton charges  are given by
\begin{eqnarray}
\label{Qmhat}
 Q_{\hat m} &=& F_A \left[ 0.093 -\frac{0.036}{A^{1/3}}- 0.020 \frac{(A-2Z)^2}{A^2} \right. \nonumber \\
  &~&~~~~\left.~~
- 1.4 \times 10^{-4} \, \frac{Z(Z-1)}{A^{4/3}} \right]  ,
\end{eqnarray}
\begin{equation}
\label{Qdeltam}
Q_{\delta m} = F_A \left[0.0017   \, \frac{A-2Z}{A} \right] ,
\end{equation}
\begin{equation}
\label{Qme}
 Q_{m_e} = F_A \left[ 5.5 \times 10^{-4}  \, \frac{Z}{A} \right] ,
\end{equation}
and
\begin{equation}
\label{Qe}
Q_e = F_A   \left[  -1.4 + 8.2 \frac{Z}{A} + 7.7 \frac{Z(Z-1)}{A^{4/3}}  \right]\times 10^{-4}.
\end{equation}
Again note that $F_A$ (see Eq. (\ref{FA})) can be readily approximated as unity.

Inspection of our general results of the previous paragraph reveals that many terms have small effects. Indeed, insertion of representative values of $A,~Z$ indicate that there are two dominant effects - those of the $A^{-1/3}$ and $Z(Z-1)/A^{4/3}$ variation in the binding energy, from strong and electromagnetic effects respectively. Until such fortunate time when we need high precision to account for multiple measurements, we can obtain a more useful formula by truncating to these two terms. Doing so yields a simpler, approximate formula involving only two dilaton charges
\begin{equation}
\label{approxalphaA}
{\alpha}_A \simeq d_g^* + \left[ (d_{\hat m} - d_g) Q'_{\hat m}  + d_e Q'_e \right]_A
\end{equation}
where
\begin{equation}
 d_g^* = d_g + 0.093 (d_{\hat m} - d_g) + 0.00027 d_e
\end{equation}
and where
\begin{equation}
\label{Qmprime}
 Q'_{\hat m} = -\frac{0.036}{A^{1/3}} - 1.4 \times 10^{-4} \, \frac{Z(Z-1)}{A^{4/3}}
\end{equation}
and
\begin{equation}
\label{Qeprime}
 Q'_{e} =  + 7.7 \times 10^{-4} \frac{Z(Z-1)}{A^{4/3}} .
\end{equation}
We think that these approximate expressions capture all the potentially dominant EP violation effects. These dilaton charges for many materials are shown in Table I.

\begin{table}[h]\centering
\caption{Approximate EP-violating `dilaton charges' for a sample of materials. These charges are averaged over the (isotopic or chemical, for SiO$_2$) composition.}
\begin{tabular}{ccccc}
\\
{\rm Material} &$A$ &$Z$ &$-Q'_{\hat m}$ &$Q'_e$ \\ \\
{\rm Li} &7 &3 &18.88 $\times 10^{-3}$ &0.345 $\times 10^{-3}$ \\
{\rm Be} &9 &4 &17.40 $\times 10^{-3}$ &0.494 $\times 10^{-3}$ \\
{\rm Al} &27 &13 &12.27 $\times 10^{-3}$ &1.48 $\times 10^{-3}$ \\
{\rm Si} &28.1 &14 &12.1 $\times 10^{-3}$ &1.64 $\times 10^{-3}$ \\
{\rm SiO$_2$} &... &... &13.39 $\times 10^{-3}$ &1.34 $\times 10^{-3}$ \\
{\rm Ti} &47.9 &22 &10.28 $\times 10^{-3}$ &2.04 $\times 10^{-3}$ \\
{\rm Fe} &56 &26 &9.83 $\times 10^{-3}$ &2.34 $\times 10^{-3}$ \\
{\rm Cu} &63.6 &29 &9.47 $\times 10^{-3}$ &2.46 $\times 10^{-3}$ \\
{\rm Cs} &133 &55 &7.67 $\times 10^{-3}$ &3.37 $\times 10^{-3}$ \\
{\rm Pt} &195.1 &78 &6.95 $\times 10^{-3}$ &4.09 $\times 10^{-3}$ \\
\end{tabular}
\end{table}

The signals for EP-violation then has two terms, namely
\begin{equation}
\left( \frac{\Delta a}{a} \right)_{BC} = (\alpha_B- \alpha_C)\alpha_E =  \left[D_{\hat m} Q'_{\hat m} + D_e Q'_e \right]_{BC}
\end{equation}
where $[Q]_{BC} \equiv Q_B - Q_C$, and where the ` dilaton charges´' are (approximately) given by Eq. (\ref{Qmprime}) and
Eq. (\ref{Qeprime}). The corresponding dilaton coefficients $D_i$ are given by
\begin{equation}
 D_{\hat m} = d^*_g \, (d_{\hat m} - d_g) \, , \qquad D_e = d^*_g \, d_e \, .
\end{equation}
If we were assuming that the dilaton parameter $d_e$ is much smaller than $d_{\hat m} - d_g$, we could go further
and conclude (in view of the numerical results indicated in Table I) that the signal $Q'_e$ is sub-dominant
w.r.t.  $Q'_{\hat m}$. In that case we would end up with a uni-dimensional EP signal proportional to $[Q'_{\hat m}]_{BC}$.

As an application of our results, let us extract the constraints on the dilaton parameters from the most sensitive present experiments. The E\"{o}tWash experiment \cite{Schlamminger:2007ht} compares Be (A=9, Z=4) and Ti (A=47.9, Z=22), with the constraint
\begin{equation}
\left(\frac{\Delta a}{a}\right)_{\rm Be \, Ti} = (\alpha_{Be}-\alpha_{Ti})\alpha_{\rm Earth} = (0.3 \pm 1.8)\times 10^{-13}
\end{equation}
Working at the two-sigma level, i.e. $(0.3 \pm 3.6)\times 10^{-13}$, and neglecting the central value $0.3$,
the rewriting of this equation in terms of the theoretical dilaton parameters  $D_{\hat m}, D_e$ yields
\begin{equation}
 | D_{\hat m} + 0.22 D_e | \le 5.1 \times 10^{-11}
\end{equation}
The Lunar Laser Ranging experiment \cite{lunarlaser} compares the acceleration of the Earth and the Moon towards the Sun, with the constraint
\begin{equation}
\left( \frac{\Delta a}{a}\right)_{\rm LLR} = (\alpha_{\rm Earth}-\alpha_{\rm Moon})\alpha_{\rm Sun} = ( -1.0 \pm 1.4)\times 10^{-13}
\end{equation}
We approximate the Earth's mantle composition as being SiO$_2$, and the Earth'core as being iron, with the Moon being similar to the mantle. The constraint here is
\begin{equation}
 | D_{\hat m} + 0.28 D_e | \le 24.6 \times 10^{-11}
\end{equation}
We see that although the Lunar experiment has a slightly better differential-acceleration sensitivity, the laboratory-based test is more sensitive to the dilaton coefficients because of a greater difference in the dilaton charges of the materials used, and of the fact that only one-third of the Earth mass is made of a different material. Note that, in the dilaton models
considered here, the EP-violation associated to the gravitational self energy \cite{Nordtvedt:1968zz} is negligible compared to the the matter couplings that we have calculated
above.

In summary we have given a general description of EP violations associated to
the possible couplings of light scalars, and in particular have calculated,
for the first time, the effects of nuclear binding (using recent work on its
quark-mass dependence). The resulting $A^{-1/3}$ dependence is seen to be one of the major factors for EP violations due to the dilaton couplings, along with the previously quantified $Z(Z-1)/A^{4/3}$ dependence of the electromagnetic interaction.
We expect that our results will be useful in assessing the optimal choices for materials in future experiments, and in interpreting their results. These upcoming experiments have the potential to probe new territory in fundamental physics and may have the opportunity to uncover rare and exciting physics.

\section*{Acknowledgements}
JFD thanks the IHES for hospitality both at the start of this project and at its conclusion. He also acknowledges support partially by the NSF grants PHY- 055304 and PHY - 0855119, and in part by the Foundational Questions
Institute. We thank Ulf Mei{\ss}ner for a useful correspondence.

\end{document}